\documentclass[reprint, amsmath, amssymb, groupaddress, aps, prb]{revtex4-1}
\usepackage{graphicx}
\usepackage{amsmath}
\begin{document}
\title{Dimension Engineering of Single-Layer PtN$_2$ with the Cairo Tessellation}
\author{Lei Liu$^{1,\dagger}$, Duo Wang$^{1,\dagger}$, Sreeharsha Lakamsani$^{2,1}$, Wenjiang Huang$^1$, Chance Price$^1$, and Houlong L. Zhuang$^{1,*}$}
\affiliation{$^1$School for Engineering of Matter Transport and Energy, Arizona State University, Tempe, AZ 85287, USA}
\email{zhuanghl@asu.edu\\$^{\dagger}$L. Liu and D. Wang contribute equally to this work.}
\affiliation{$^2$Hamilton High School, Chandler, AZ 85248, USA}
\date{\today}
%----------------------------------------------------------------------
\begin{abstract}
Single-layer PtN$_2$ exhibits an intriguing structure consisting of a tessellation pattern called the Cairo tessellation of type 2 pentagons, which belong to one of the existing 15 types of convex pentagons discovered so far that can monohedrally tile a plane. Single-layer PtN$_2$ has also been predicted to show semiconducting behavior with direct band gaps. Full exploration of the structure-property relationship awaits the successful exfoliation or synthesis of this novel single-layer material, which depends on the structure of its bulk counterpart with the same stoichiometry to some extent. Bulk PtN$_2$ with the pyrite structure is commonly regarded as the most stable structure in the literature. But comparing the energies of single-layer PtN$_2$ and bulk PtN$_2$ leads to a dilemma that a single-layer material is more stable than its bulk counterpart. To solve this dilemma, we propose stacking single-layer PtN$_2$ sheets infinitely to form a new bulk structure of PtN$_2$. The resulting tetrahedral layered structure is energetically more stable than the pyrite structure and single-layer PtN$_2$. We also find that the predicted bulk structure is metallic, in contrast to the semiconducting pyrite structure. In addition to predicting the 3D structure, we explore the possibility of rolling single-layer PtN$_2$ sheets into nanotubes. The required energies are comparable to those needed to form carbon or boron nitride nanotubes from their single-layer sheets, implying the feasibility of obtaining PtN$_2$ nanotubes. We finally study the electronic structures of PtN$_2$ nanotubes and find that the band gaps of PtN$_2$ nanotubes are tunable by changing the number of unit cells $N$ of single-layer PtN$_2$ used to construct the nanotubes. Our work shows that dimension engineering of PtN$_2$ not only leads to a more stable 3D structure but also 1D materials with novel properties. 
\end{abstract}
\maketitle
%----------------------------------------------------------------------
\section{Introduction}
A number of two-dimensional (2D) materials have been predicted and recorded in various databases such as the computational 2D materials database (C2DB)\cite{haastrup2018computational} and Materialsweb.\cite{ashton2017topology} But many of these materials, in spite of their exotic properties, exhibit no known bulk counterparts especially those with the same stoichiometry, making it challenging to obtain these 2D materials. Being such an example, single-layer platinum nitride PtN$_2$ has recently been predicted in several theoretical studies.\cite{liu2018penta,zhao20192d,yuan2019single} The reason for the uniqueness of this single-layer material is twofold: First, the structure as illustrated in Fig.\ref{fig:structures}(a) is completely planar with a tessellation of type 2 pentagons that are able to tessellate a plane; This tessellation is called the Cairo tessellation, as it appears in the streets of Cairo.\cite{wells1991penguin} Second, it is a semiconductor with predicted high carrier mobility and Young's modulus.\cite{liu2018penta} The peculiar structure and properties of single-layer PtN$_2$ call for its synthesis, which largely rely on the existence of stable structure of bulk PtN$_2$.

Even without in the above context of single-layer PtN$_2$, bulk PtN$_2$ on its own has attracted considerable attention as an example in the family of transition-metal nitrides, which generally possess notable electrical, mechanical, and thermal properties.\cite{yamanaka1998s,young2006interstitial,fu2009theoretical}  
The pyrite structure of bulk PtN$_2$ is commonly regarded as the most stable.\cite{yu2006elastic,gou2006theoretical,crowhurst2006synthesis,young2006interstitial} Figure~\ref{fig:structures}(b) illustrates the pyrite structure, consisting of Pt atoms occupying the lattice sites of a face-centered cubic (fcc) lattice and each Pt atom is six-fold coordinated with N atoms to form corner-sharing Pt-N octahedra. Two other possible structures including the fluorite (as shown in Fig.\ref{fig:structures}(c)) and marcasite (as shown in Fig.\ref{fig:structures}(d)) structures have also been studied.\cite{yu2005platinum,yu2006elastic,chen2007crystal} In the fluorite structure, Pt atoms are also located at the fcc lattice sites, but each Pt atom has eight nearest neighboring N atoms. In the marcasite structure, the Pt and N atoms also form corner-sharing Pt-N octahedra, but the Pt atoms occupy the sites of a body-centered tetragonal lattice.

In addition to 2D materials, 1D nanotubes have sparked wide interest since the discovery of carbon nanotubes (CNTs).\cite{iijima1991helical,bacsa1995aligned} Their mechanical, electrical, and optical properties can be tuned by modifying the diameters and charality,\cite{hamada1992new} making CNTs promising for a wealth of applications,\cite{koziol2007high} such as field emission electron source \cite{de1995carbon} and light-emitting diodes.\cite{chen2005bright} Successful fabrication of CNTs indicates the feasibility of obtaining non-carbon nanotubes based on other single-layer materials. Indeed, extensive experimental and theoretical research has been extended to study boron nitride BN, carbonitrides B$_x$C$_y$N$_z$, and transition-metal dichalcogenides $MX_2$ ($M$ and $X$ represent transition-metal and chalcogen elements, respectively) nanotubes.\cite{zettl1996non,pokropivny2001non, Ivanovskii_2002} 

Although many 2D materials and their structures have been predicted based on the bulk counterparts of these 2D materials with the same stoichiometry, we revert this process in this work by first showing a counter-intuitive result that single-layer PtN$_2$ is more energetically stable than bulk PtN$_2$ with the pyrite structure. We then study the interactions between two layers of PtN$_2$ and suggest a layered structure with on-top stacking of single-layer PtN$_2$ sheets as the more stable bulk structure. Furthermore, due to the above-mentioned excellent properties of nanotubes, we explore the structure-property relationships of PtN$_2$ nanotubes.
%----------------------------------------------------------------------
\section{Methods}
All the DFT calculations are performed using the Vienna {\it Ab-initio} Simulation Package (VASP, version 5.4.4)\cite{Kresse96p11169} We apply the Perdew-Burke-Ernzerhof (PBE) functional to approximate the exchange-correlation interactions.\cite{Perdew96p3865} We use Grimme's DFT-D3 method to describe the van der Waals (vdW) interactions in bilayer PtN$_2$ and our proposed layered structure of bulk PtN$_2$.\cite{grimme2010consistent} We also use the optB88-vdW functional to compare against the accuracy of some of the DFT-D3 results.\cite{dion2004van,roman2009efficient,klimevs2011van} We use the standard potential datasets created with the PBE functional for Pt and N generated according to the projector augmented wave method .\cite{Bloechl94p17953,Kresse99p1758} These datasets treat the 5$d^9$ and 6$s$ electrons of Pt atoms and the 2$s^2$ and 2$p^3$ electrons of N atoms as valence electrons. Plane waves with their cut-off kinetic energies below 550 eV are used to approximate the electron wave functions. We use a $\Gamma$-centered $12~\times~12~\times~1$ Monkhorst-Pack\cite{PhysRevB.13.5188} $k$-point grid for single-layer and bilayer PtN$_2$, and a $12~\times~12~\times~12$ grid for bulk PtN$_2$ with the pyrite, fluorite, marcasite, and AB-stacked structures, and a $12~\times~12~\times~15$ grid for bulk PtN$_2$ with the tetragonal AA-stacked layered structure, and a $9~\times~1~\times~1$ grid for PtN$_2$, carbon, and boron nitride nanotubes, to sample the $k$ points in the reciprocal space. A sufficiently large vacuum spacing ($>$ 18.0~\AA) is applied to the slabs of single-layer and bilayer PtN$_2$ and nanotubes to avoid the image interactions due to the periodic boundary conditions. The lattice constants and atomic coordinates of bulk PtN$_2$ with different structures are completely optimized. For single-layer and bilayer PtN$_2$, we optimize the in-plane lattice constant and the atomic positions. For the PtN$_2$ nanotubes, we relax only the lattice constant along the tube direction and the atomic positions. The force threshold value for all of these geometry optimizations is the same, i.e., 0.01 eV/\AA. 
%----------------------------------------------------------------------
\section{Results and Discussion} 
\begin{figure}
 \includegraphics[width=8cm]{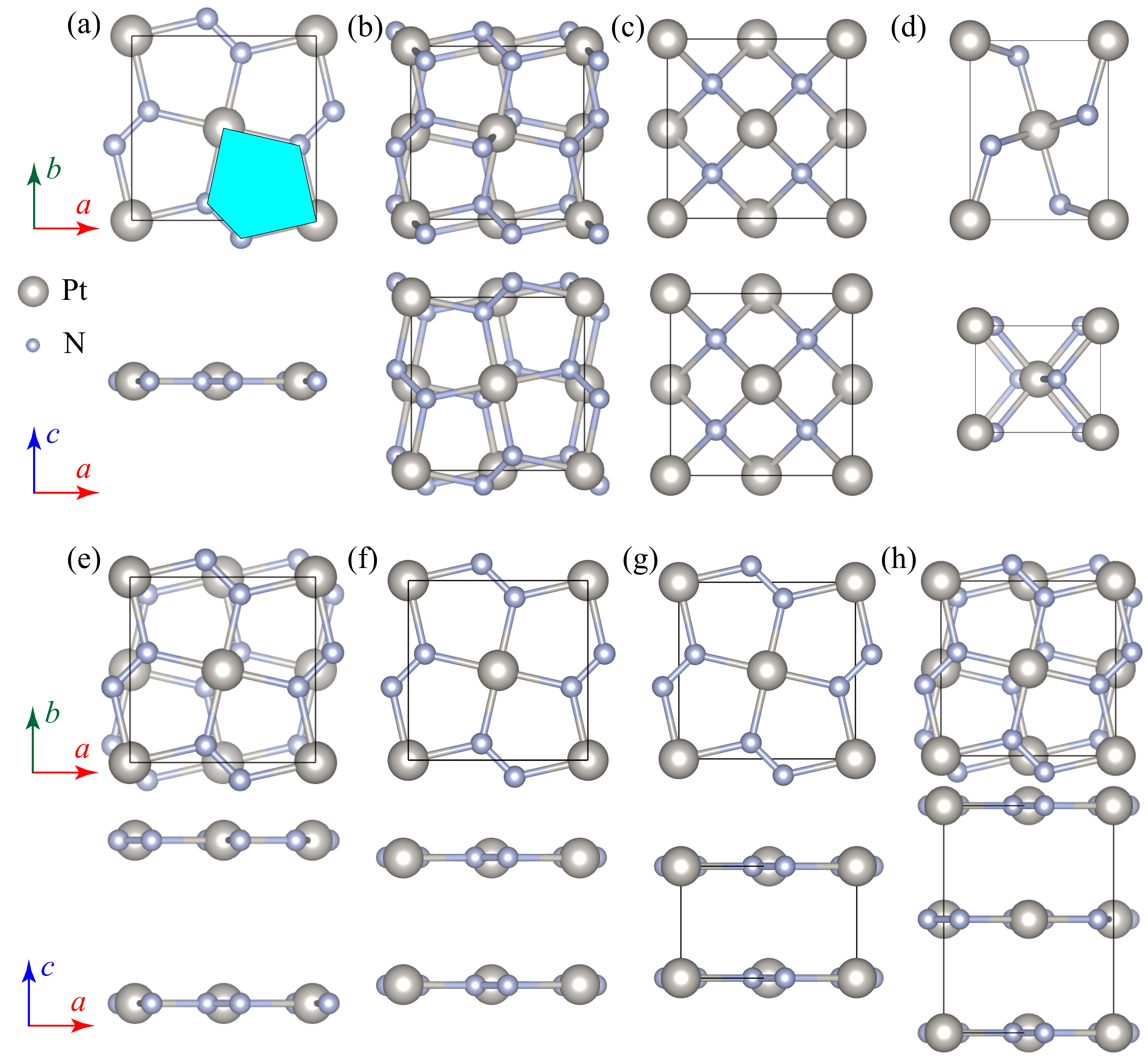}
 \caption{Top and side views of the unit cells of (a) single-layer PtN$_2$ and bulk PtN$_2$ with the (b) pyrite, (c) fluorite, and (d) marcasite structures, and of (e) AB-stacked and (f) AA-stacked bilayer PtN$_2$, and of bulk PtN$_2$ with (g) AA-stacked and (h) AB-stacked tetragonal layered structures. A type 2 pentagon is enclosed by the cyan shaded area sketched in (a).}
	\label{fig:structures}
\end{figure}
\begin{figure}
\includegraphics[width=8cm]{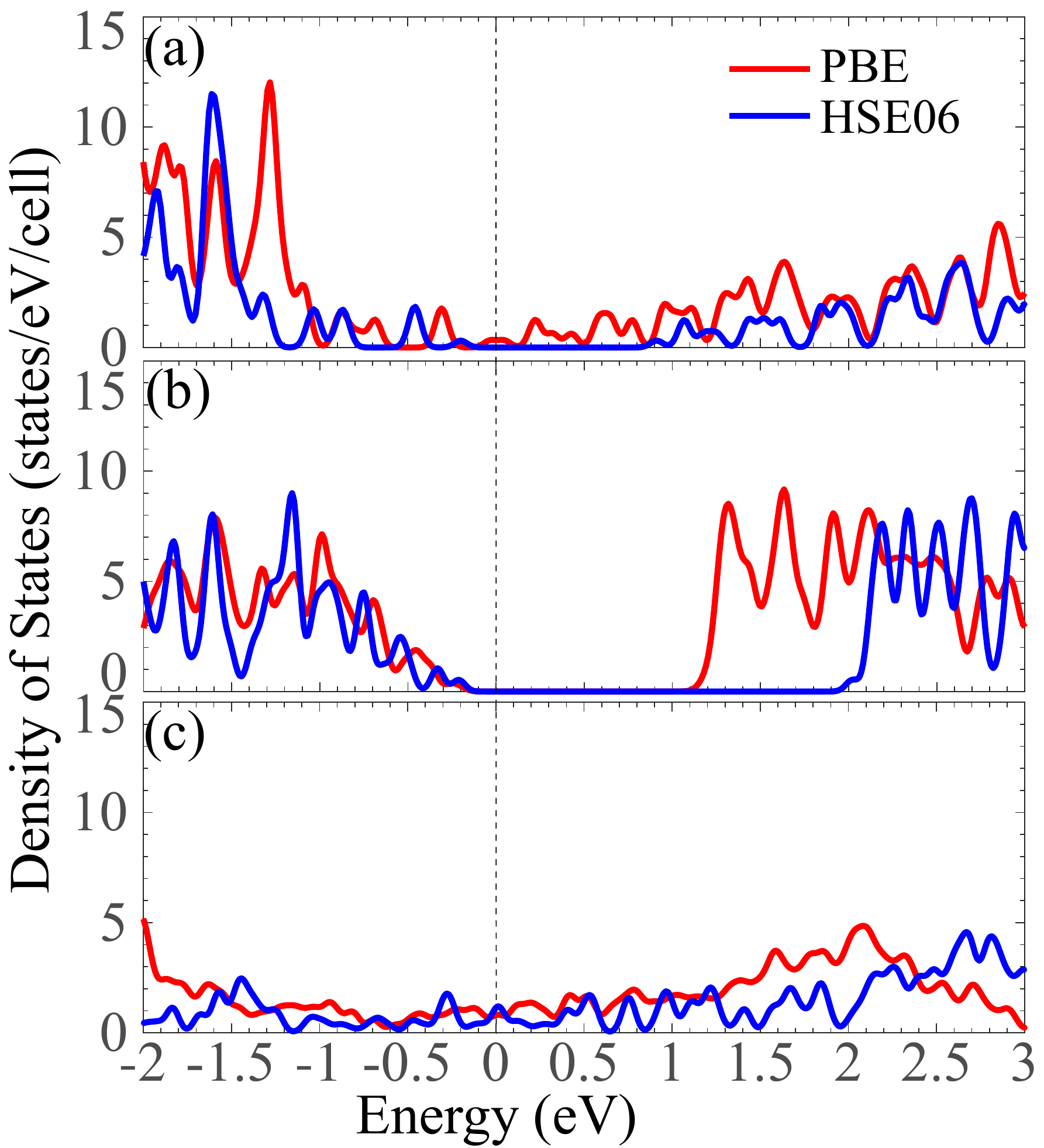}
\caption{Density of states of (a) single-layer PtN$_2$, (b) bulk PtN$_2$ with the pyrite structure, and (c) bulk PtN$_2$ with the tetragonal layered structure calculated with the PBE and HSE06 functionals.}
\label{fig:dos}
\end{figure}

We first benchmark our calculations on single-layer PtN$_2$ with previous theoretical studies. Our calculated in-plane lattice constant (4.81\AA) is consistent with the reported results (4.80,\cite{liu2018penta} 4.81,\cite{yuan2019single} and 4.83\cite{zhao20192d}~\AA). For the electronic structure, Fig.~\ref{fig:dos}(a) shows the density of states (DOS) of single-layer PtN$_2$ calculated with the PBE and HSE06 functionals. The PBE functional seriously underestimates the band gap of single-layer PtN$_2$. The PBE DOS curve shows that this functional actually leads to a conclusion that single-layer PtN$_2$ is metallic, agreeing with the rather small bandgaps (0.075 and 0.07 eV) reported in Refs. [\onlinecite{zhao20192d}] and [\onlinecite{yuan2019single}], respectively. Our HSE06 DOS shows a corrected band gap of single-layer PtN$_2$ as 1.11 eV, which is the same as the band gaps reported in Refs.[\onlinecite{liu2018penta}] and [\onlinecite{yuan2019single}]. Note that the work of Yang {\it et al} also considers spin-orbit coupling (SOC) and the PBE+SOC and HSE06+SOC band gaps (0.33 and 1.17 eV) are slightly larger than the PBE and HSE06 band gaps.\cite{liu2018penta} 

We next calculate the energy difference between single-layer PtN$_2$ and bulk PtN$_2$ with the pyrite, fluorite, and marcasite structures. This energy difference (i.e., $E_\mathrm{2D}$-$E_\mathrm{3D}$, the formation energy of 2D materials) is a  metric of the energy cost to exfoliate a single-layer nanosheet from its 3D counterpart and also an indicator of the feasibility of chemical synthesis.\cite{singh2015computational} We find the formation energies with reference to the three bulk structures are all negative: -168, -1076, and -207 meV/atom. The more negative formation energy implies the less stable of the bulk structure used for comparison. These energy differences therefore show that the pyrite structure is the most stable in comparison with the fluoride and marcasite structures, consistent with previous theoretical studies.\cite{chen2010ab, crowhurst2006synthesis,yu2006elastic} We conclude that single-layer PtN$_2$ is more stable than the pyrite structure. More important, the negative formation energies show that all the three bulk structures used for references are not the ground state of bulk PtN$_2$. In parallel with this observation, we also perform the same calculations on single-layer graphene and compare its energy to the face-centered-cubic diamond structure. We obtain a negative formation energy of -128 meV/atom using the PBE functional, which is expected as the bulk ground state is graphite.

\begin{table*}
\caption{Relative energy (in meV/atom) of single-layer PtN$_2$, bulk PtN$_2$ with the pyrite, fluorite, and marcasite structures, AB and AA-stacked bilayer PtN$_2$, and AB-stacked bulk PtN$_2$. The energy of the tetragonal AA-stacked layered structure is set to zero. All the relative energies are calculated using both the PBE and DFT-D3 methods.}
\begin{ruledtabular}
\begin{center}
\begin{tabular}{cccccccc}
Method & Single-layer & Bulk-pyrite & Bulk-fluorite & Bulk-marcasite  & Bilayer-AB & Bilayer-AA & Bulk-AB \\
\hline
PBE& 34  & 202 & 1110 & 241  & 33 & 25 & 97 \\
DFT-D3 & 199 & 139 & 1067 &184  & 156 & 130 & 156\\
\end{tabular}
\end{center}
\end{ruledtabular}
\label{summary}
\end{table*}

\begin{figure}
     \includegraphics[width=8cm]{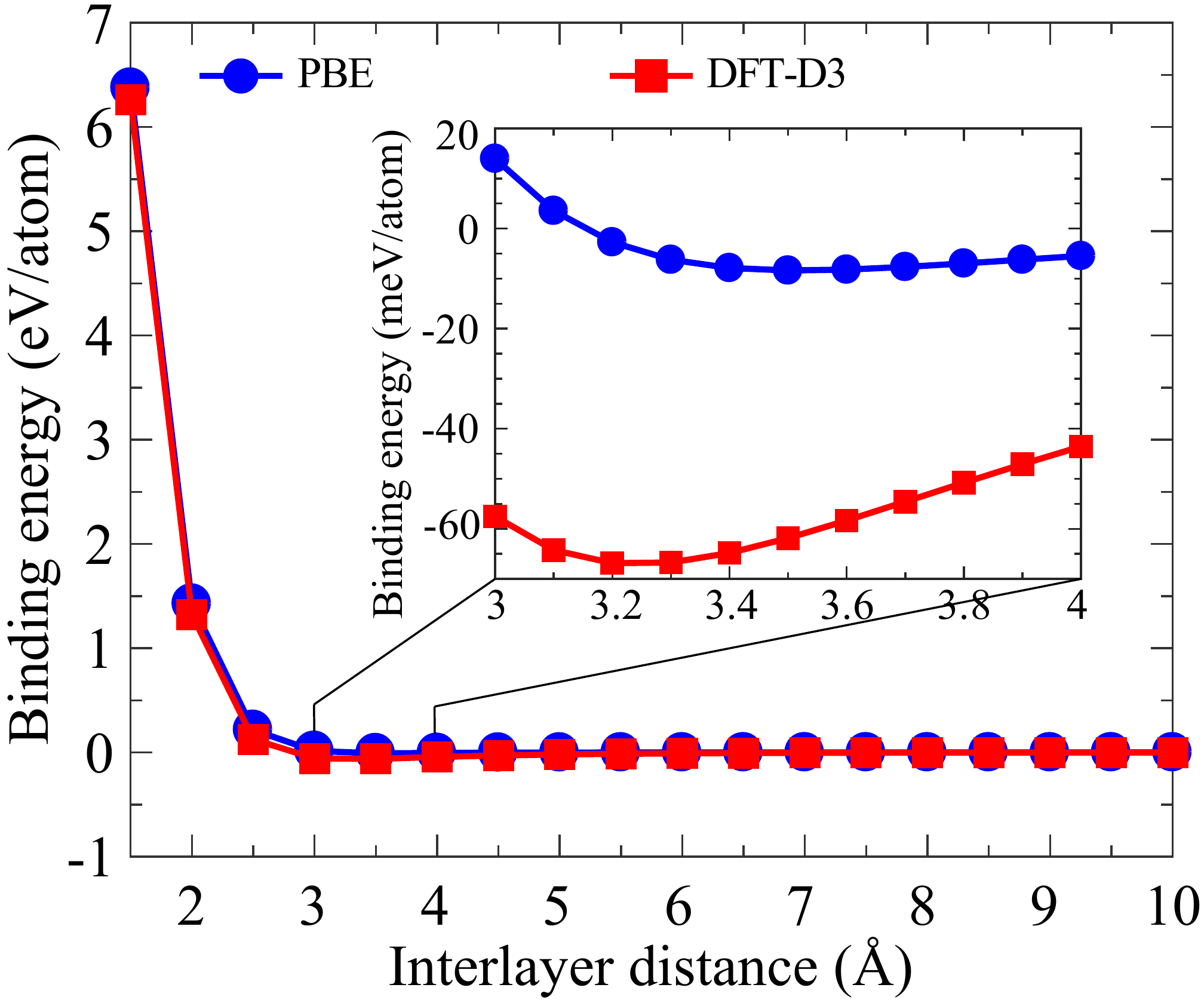}
	\caption{Binding energy of AA-stacked bilayer PtN$_2$ as a function of the interlayer distance computed using the PBE and DFT-D3 methods. The inset is an enlarged view of the two binding energy curves where interlayer distance ranges from 3 to 4~\AA.}
	\label{fig:binding}
\end{figure}

To search for the more stable bulk structure, we begin with studying the energy change by stacking two sheets of single-layer PtN$_2$ to form bilayer PtN$_2$. We account for two types of stacking for bilayer PtN$_2$. One is called the AB stacking (see Fig.~\ref{fig:structures}(e)), where the Pt atoms in one layer of bilayer PtN$_2$ are located above/below the center of a pair of N atoms in another layer. The other one is the AA stacking (see Fig.~\ref{fig:structures}(f)), where the second layer is located on top of the first layer. We find that AB-stacked bilayer PtN$_2$ is energetically less stable than the AA-stacked structure by 8 meV/atom, so we focus on the AA-stacked bilayer PtN$_2$ and compute the binding energy $E_b$ between the two layers defined as $E_b$ = $E_\mathrm{bilayer}$-2$E_\mathrm{single-layer}$. Figure~\ref{fig:binding} displays the $E_b$ of AA-stacked bilayer PtN$_2$ as a function of the interlayer distance. As can be seen, without using the DFT-D3 method to describe the vdW interactions, the $E_b$ values resulting from the interactions between the two PtN$_2$ layers are negligibly small with the maximum binding energy of -9 meV/atom. Taking into account the vdW interactions, the binding energy is corrected to -69 meV/atom (we obtain the same binding energy using the optB88-vdW functional), which is similar to the binding energy (-31.1 meV/atom) of AA-stacked bilayer graphene calculated using the DFT-D method,\cite{PhysRevLett.115.115501} showing the weak interactions between single-layer PtN$_2$ sheets. 

Because bilayer AA-stacked PtN$_2$ is more stable than single-layer PtN$_2$, we expect to stack an infinite number of single-layer PtN$_2$ sheets in the AA-stacking manner to result in a more stable structure of bulk PtN$_2$ as illustrated in Fig.~\ref{fig:structures}(g). A symmetry analysis of this infinitely AA-stacked layered structure shows that the bulk structure is tetragonal with the space group P4/mbm (No.127). We henceforth refer to this new bulk structure as the tetragonal layered (TL) structure. For the completeness of comparison, we also compute the energy of the bulk structure with AB-stacked layers (see Fig.~\ref{fig:structures}(h)). We find that the TL structure is the most stable among the eight structures displayed in Fig.~\ref{fig:structures}. Table~\ref{summary} lists all the relative energies of the eight structures using the energy of the TL structure as the reference. The lattice constants of these two bulk structures are reported in Table~\ref{summary2}. With the new bulk structure, the formation energies of single-layer PtN$_2$ become physically positive, which are 34 and 199 meV/atom calculated with the PBE and DFT-D3 methods, respectively. These small formation energies manifest the weak interactions between layers and also indicate a feasible approach to obtain single-layer may be the mechanical exfoliation method as used to obtain single-layer graphene.\cite{novoselov2004electric}

Since we have identified the more energetically stable structure of bulk PtN$_2$, we now examine the mechanical stability of bulk PtN$_2$ with the TL structure. We also calculate the same properties of bulk PtN$_2$ with the pyrite structure for comparison, as the pyrite structure is the most stable among the previously reported bulk structures. Table~\ref{summary2} summarizes the predicted independent elastic stiffness constants for cubic and tetragonal PtN$_2$ using a symmetry-general approach.\cite{PhysRevB.65.104104}. According to Born's criteria of mechanical stability, the following conditions:\cite{PhysRevB.90.224104}
 \begin{equation}
C_{11} - C_{12} > 0, C_{11} + 2 C_{12} > 0, C_{44} > 0
\label{eq1}
\end{equation}
and
\begin{equation}
\begin{split}
&C_{11} > |C_{12}|, 2C_{13}^2 < C_{33}(C_{11} + C_{12}),\\
&C_{44} > 0, C_{66} > 0
\label{eq2}
\end{split}
\end{equation}
 need to be satisfied for cubic and tetragonal PtN$_2$, respectively. The computed elastic constants show that both the pyrite and TL structures of bulk PtN$_2$ are mechanically stable.

\begin{table*}
\caption{Lattice constants (in \AA) and elastic stiffness constants and hardness (in GPa) of bulk PtN$_2$ with the pyrite and tetragonal layered (TL) structures. Theoretical results using different methods are cited for comparison. For the pyrite structure, $C_{13}$ = $C_{12}$, $C_{33}$ = $C_{11}$, and $C_{66}$ = $C_{44}$ due to the cubic symmetry.}
\begin{ruledtabular}
\begin{center}
\begin{tabular}{cccccccccccccc}
Structure &Method & $a$ & $b$ & $c$ & $C_{11}$ & $C_{12}$ & $C_{13}$ & $C_{33}$ & $C_{44}$ & $C_{66}$ & $B_\mathrm{VRH}$ & $G_\mathrm{VRH}$ & $H$\\
\hline
Pyrite & PBE & 4.85 & 4.85 & 4.85 & 695 & 87 & - & - & 133 & - & 290 & 187 & 22.5\\
       &DFT-D3 & 4.82 & 4.82 & 4.82 & 746 & 93  & - & - & 136 & - &311 & 195  & 22.3\\
       & LDA-NC\footnote{Ref.[\onlinecite{garcia2018ab}]; NC: norm-conserving pseudopotentials} & 4.81 & 4.81 & 4.81 & 828 & 113 & - & - & 155 & - & 351 & 218 & 23.7, 27.7\footnote{Elastic stiffness constants are calculated from only the Voigt approximation.}\\
       & PW91-PAW\footnote{Ref.[\onlinecite{liu2014electronic}]} & 4.88 & 4.88 & 4.88 & 662 & 69 & - & - & 129 & - & 267 & 181 & 23.7\\
       & LDA-LAPW\footnote{Ref.[\onlinecite{yu2006elastic}]; LAPW: linearized augmented plane waves} & 4.77 & 4.77 & 4.77 & 824 & 117 & - & - & 152 & - & 353 & 215 & 22.9 \\
       &PBE-LAPW$^\mathrm{d}$ & 4.86 & 4.86 & 4.86 & 668 & 78 & - & - & 133 & - & 275 & 184 & 23.5 \\
TL     &PBE  & 4.83 & 4.83 & 3.07 & 709 & 120 & 18 & 55 & 14 & 135 & 125 & 74 & 10.5\\
       &DFT-D3  & 4.81 & 4.81 & 2.90 & 782 & 134 & 17 & 110 & 18 & 143 & 159 & 87 & 10.4\\
\end{tabular}
\end{center}
\end{ruledtabular}
\label{summary2}
\end{table*}

Hardness is an important property of platinum nitrides for their engineering applications.\cite{chen2005hard} We therefore calculate the Vicker hardness $H_\mathrm{V}$ of bulk PtN$_2$ with the pyrite and TL structures using the following empirical equation:\cite{chen2011modeling}
\begin{equation}
H_\mathrm{V} = 2 (G_\mathrm{VRH}^3/B_\mathrm{VRH}^2)^{0.585} - 3,
\label{eq5}
\end{equation}
where the bulk and shear moduli ($B_\mathrm{VRH}$ and $G_\mathrm{VRH}$) are calculated using the Voigt?-Reuss-?Hill (VRH) approximation:\cite{voigt1928lehrbuch,hill1952elastic} 
\begin{equation}
B_\mathrm{VRH} = \frac{B_\mathrm{V}+B_\mathrm{R}}{2}
\label{eq55}
\end{equation}
and
\begin{equation}
G_\mathrm{VRH} = \frac{G_\mathrm{V}+G_\mathrm{R}}{2}.
\label{eq55}
\end{equation}
$B_\mathrm{V}$ and $G_\mathrm{V}$ are the upper bounds of bulk and shear moduli written as
\begin{equation}
B_\mathrm{V} = \frac{2 C_{11} + 2 C_{12} + C_{33} + 4 C_{13}}{9}\\
\label{eq33}
\end{equation}
and
\begin{equation}
G_\mathrm{V} = \frac{2 C_{11} - C_{12} + C_{33} - 2 C_{13} + 6 C_{44} + 3 C_{66}}{15},
\label{eq44}
\end{equation}
respectively. $B_\mathrm{R}$ and $G_\mathrm{R}$ are the lower bounds of bulk and shear moduli, i.e., 
\begin{equation}
B_\mathrm{R} = \frac{1}{2 S_{11} + 2 S_{12} + S_{33} + 4 S_{13}}\\
\label{eq55}
\end{equation}
and
\begin{equation}
G_\mathrm{R} = \frac{15}{8 S_{11} - 4 S_{12} + 4 S_{33} - 8 S_{13} + 6 S_{44} + 3 S_{66}},
\label{eq66}
\end{equation}
where $S_{ij} (i, j = 1-6)$ are the elastic compliant constants and matrix $S$ is equal to the inverse of matrix $C$. Table~\ref{summary2} lists the predicted $B_\mathrm{VRH}$ and $G_\mathrm{VRH}$, and $H_\mathrm{V}$ obtained from using the PBE and DFE-D3 methods. Our calculated elastic constants agree well with those reported Refs.~[\onlinecite{liu2014electronic}] and [\onlinecite{yu2006elastic}] computed using the PW91 and PBE functionals with the general gradient approximation.\cite{PhysRevB.46.6671} The local-density approximations (LDA)\cite{PhysRev.140.A1133,PhysRevLett.45.566} are also used in Refs.~[\onlinecite{garcia2018ab}] and [\onlinecite{yu2006elastic}], but this method seems to lead to larger elastic constants. However, the resulting hardness (22.5 GPa) in this work is similar to those in all the references, in spite of the methods used. Although several elastic stiffness constants of the TL structure (e.g., $C_{33}$ are significantly affected by the consideration of the vdW interactions, Table~\ref{summary2} also shows that our hardness values from the PBE and DFT-D3 methods are similar. The much smaller Vicker hardness of the TL structure suggests that it is much softer than the pyrite structure. This softness appears to be a common feature for layered materials such as graphite in contrast to superhard diamond.\cite{hornyak2008introduction}

 \begin{figure}
     \includegraphics[width=8cm]{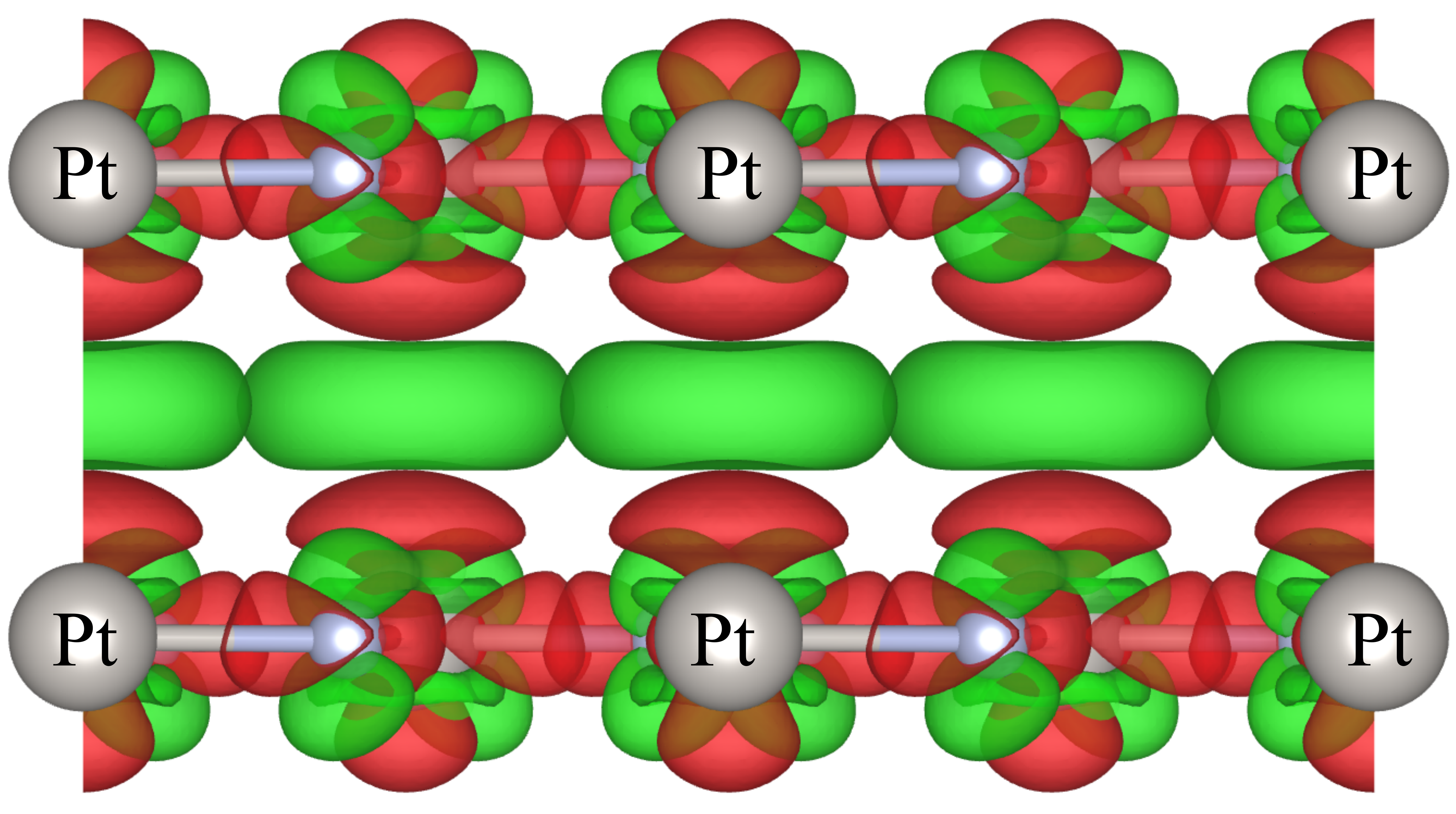}
	\caption{Charge density difference between AA-stacked bilayer PtN$_2$ and two isolated sheets of single-layer PtN$_2$. Green and red isosurfaces represent charge accumulation and depletion, respectively. The isosurface value is 2$\times10^{-4}e/a_0^3$ ($a_0$: Bohr radius).}
	\label{fig:chgd}
\end{figure}
We now compare the electronic structures of bulk PtN$_2$ with the pyrite and TL structures. Figure~\ref{fig:dos}(b) shows the density of states of bulk PtN$_2$ with the pyrite and TL structures calculated with the PBE and HSE06 functionals. For the former structure, both the PBE and HSE06 functionals predict that it is semiconducting and the band gaps are 1.35 and 2.22 eV, respectively. Our calculated PBE band gap is consistent with the previously reported band gap of 1.30 eV.\cite{young2006interstitial} For the TL structure, the PBE and HSE06 functionals consistently show that the structure is metallic. To gain a qualitative understanding of the semiconductor-to-metal transition as the number of single-layer sheets increases, we calculate the charge density difference between AA-stacked bilayer PtN$_2$ and two isolated single-layer PtN$_2$ sheets. Figure~\ref{fig:chgd} shows that the electrons of Pt atoms in both layers are transferred to the region between two Pt atoms, when the two layers interact to form bilayer PtN$_2$. These electrons between the layers form Pt-Pt metallic bonds, leading to the metallic behavior of bilayer as well as bulk PtN$_2$. In other words, the interlayer interactions in bilayer PtN$_2$ or bulk PtN$_2$ with the TL structure consist of mixed vdW and metallic bonding types. Note that the bond strength of these metallic bonds is small as reflected by the small isosurface value. We also expect the metallic bonding is significantly smaller than the vdW interactions, as including the vdW interactions drastically changes the energy difference between single-layer and bilayer PtN$_2$ (see Table \ref{summary}). Extracting a sheet of single-layer PtN$_2$ from bilayer and bulk PtN$_2$ prohibits the delocalization of the electrons. As a result, the electrons are localized around Pt atoms in the region enclosed by the red isosurface as shown in Fig.\ref{fig:chgd}, causing single-layer PtN$_2$ to be semiconducting.

\begin{figure}
     \includegraphics[width=8cm]{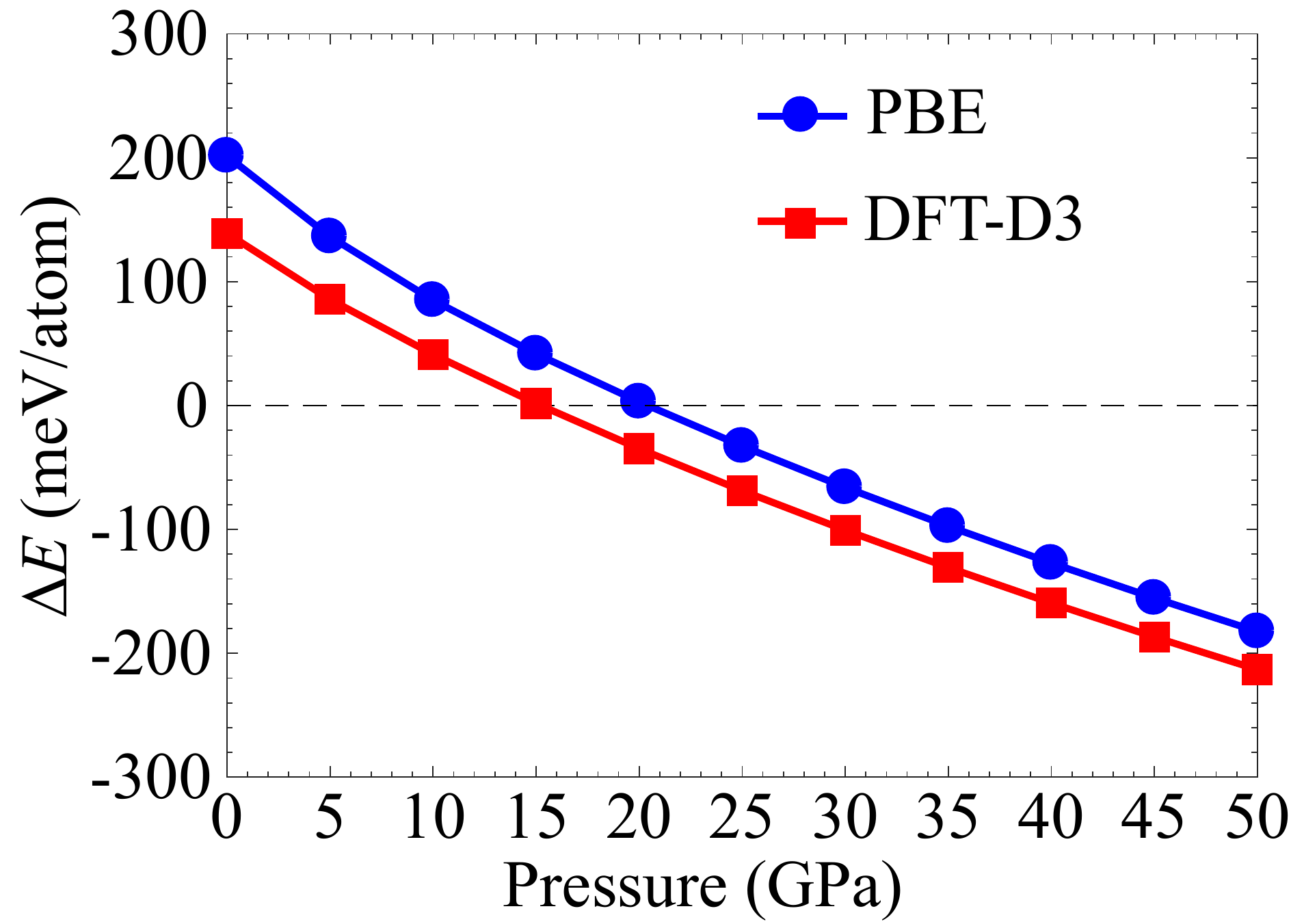}
	\caption{Pressure-dependent energy difference $\Delta E$ ($\Delta E$ = $E_\mathrm{pyrite}$-$E_\mathrm{TL}$) between bulk PtN$_2$ with the pyrite and tetragonal layered (TL) structures.}
	\label{fig:evp}
\end{figure}
\begin{figure}
     \includegraphics[width=8cm]{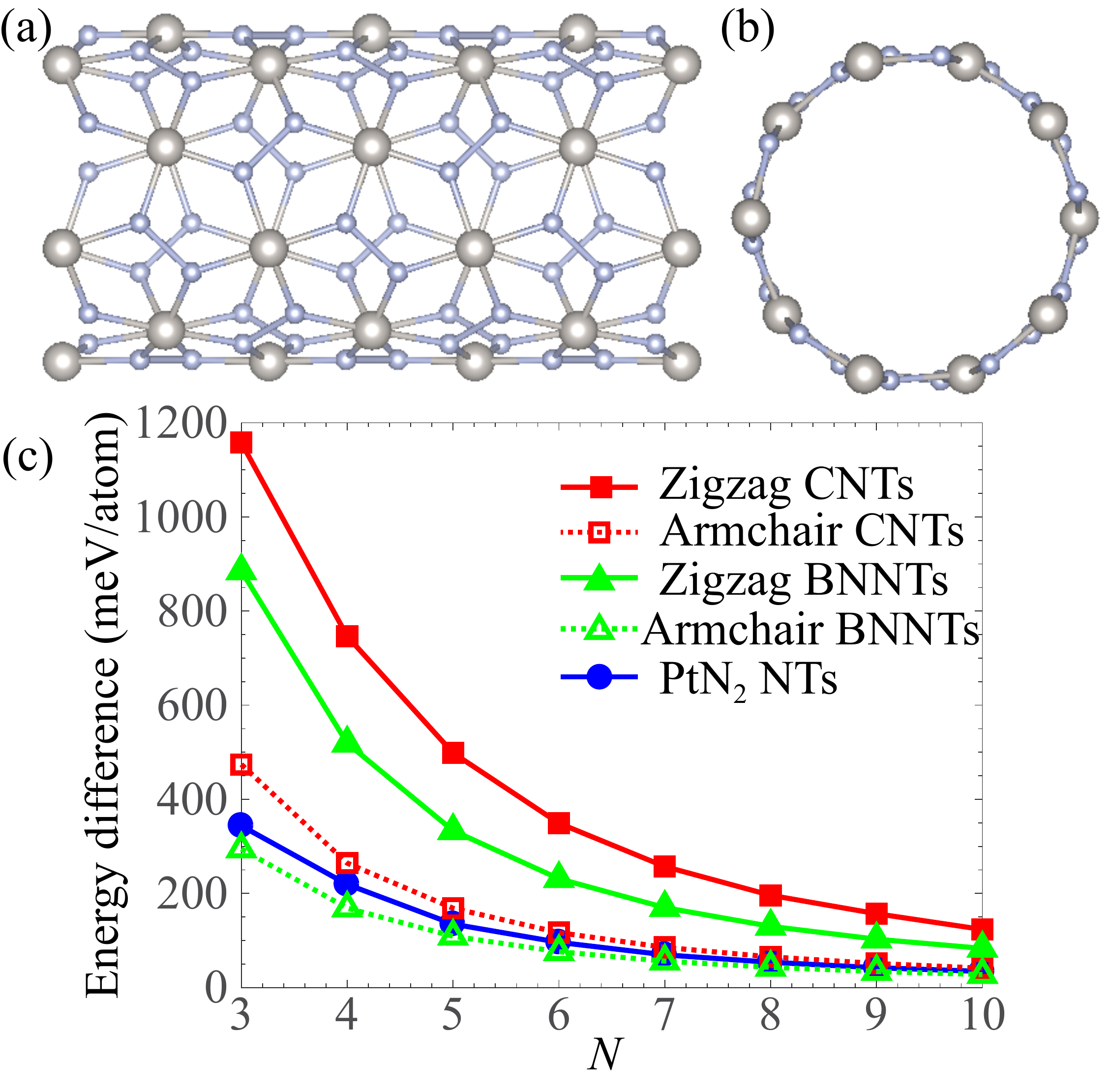}
	\caption{(a) Side and (b) top views of a PtN$_2$ nanotube model formed by wrapping a 5 $\times$ 3 $\times$ 1 supercell of single-layer PtN$_2$ about the a/b axis denoted in Fig.\ref{fig:structures}. (c) $N$-dependent energy difference between PtN$_2$ nanotubes (NTs) and single-layer PtN$_2$. The energy differences for zigzag and armchair carbon nanotubes (CNTs) and boron nitride nanotubes (BNNTs) are also shown for comparison.}
	\label{fig:nanotube}
\end{figure}
Experiments on bulk PtN$_2$ with the pyrite structure indicate the importance of stabilizing this bulk phase by external pressure.\cite{gregoryanz2004synthesis, ivanovskii2009platinum} We therefore compare the stability of bulk PtN$_2$ with the pyrite and TL structures at different pressures by computing their energy difference $\Delta E$ = $E_\mathrm{pyrite}$-$E_\mathrm{vdW}$. Figure~\ref{fig:evp} shows $\Delta E$ as a function of pressure calculated with the PBE and DFT-D3 methods. The two curves reveal the same trend: $\Delta E$ changes almost linearly from positive to negative as pressure increases, showing that the TL structure is more stable below a critical pressure, above which the pyrite structure is more stable. This trend may be caused by the exponentially increased energy as the interlayer distance in the TL structure decreases due to the pressure (see Fig.\ref{fig:binding}). We also find that the critical pressures resulted from the PBE and DFT-D3 methods are similar. For the PBE method, the transition pressure is around 20 GPa; For the latter method, the pressure is about 15 GPa. 

Having studied the case of increasing the dimension of PtN$_2$ from 2D to 3D, we set to reduce the dimension to 1D to obtain PtN$_2$ nanotubes. Many 2D materials such as single-layer graphene and boron nitride have their corresponding forms of nanotubes and exhibit novel properties.\cite{ZHI201092,chen2017quantitative} We create simulation models of PtN$_2$ nanotubes by wrapping $N$ $\times$ 1 $\times$ 1 (3 $\leq N \leq$ 10) supercells of single-layer PtN$_2$ about the $a$ axis as shown in Fig.\ref{fig:structures}(a). Due to the square symmetry of single-layer PtN$_2$, wrapping the supercells about the $b$ axis leads to the same nanotubes. The integer $N$ therefore controls the diameters of the nanotubes. Figure~\ref{fig:nanotube}(a) and (b) illustrates the side and top views of a model of PtN$_2$ nanotube. Notice that the side view actually demonstrates a curved Cairo tessellation pattern of type 2 pentagons. 

We compute the energy cost to obtain these nanotubes using the energy of single-layer PtN$_2$ as a reference. We additionally calculate the energy costs of wrapping single-layer graphene and boron nitride into zigzag ($N$, 0) and armchair ($N$, $N$) nanotubes for comparison. As can be seen from Fig.\ref{fig:nanotube}(c), the energy costs of all the nanotubes decrease with the increasing sizes of the nanotubes. This trend is expected as $N$ increases towards infinity, the diameters increase along with the decreasing curvatures of the nanotubes until they are close to zero, corresponding to the curvature of single-layer planar PtN$_2$. We observe that the energy costs of PtN$_2$ nanotubes are much smaller than those of zigzag carbon and BN nanotubes with the same $N$ values. The energy costs of CNTs and BNNTs are significantly dependent on the chirality, i.e., the energy costs of armchair CNTs and BNNTs are drastically smaller and comparable to those of armchair nanotubes. Zigzag and armchair CNTs and BNNTs have been successfully synthesized,\cite{janas2018towards} indicating that it is also possible to synthesize PtN$_2$ nanotubes. From the geometry perspective, if successfully synthesized, PtN$_2$ nanotubes will be the first nanotubes with a curved Cairo tessellation of type 2 pentagons.

\begin{figure}
     \includegraphics[width=8cm]{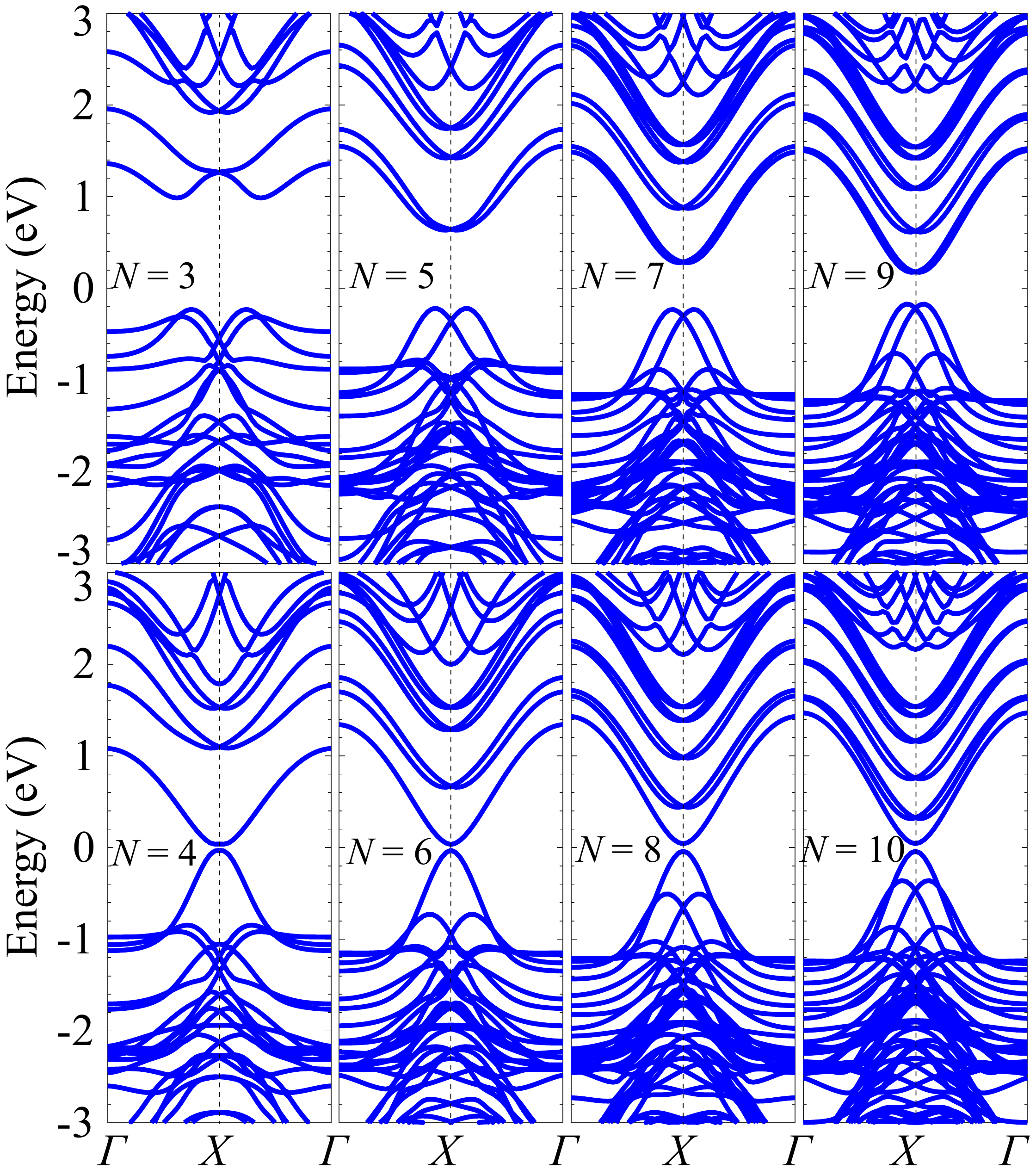}
	\caption{Band structures of PtN$_2$ nanotubes with $N$ ranging from 3 to 10 calculated at the DFT-PBE level of theory.}
	\label{fig:ntel}
\end{figure}

Finally, we calculate the electronic structures of PtN$_2$ nanotubes. Figure~\ref{fig:ntel} shows the PBE band structures of PtN$_2$ nanotubes with the eight $N$ values. We notice that these band structures strongly depend on the $N$ values, similar to the dependence of the electronic structures of CNTs on their chiral indices.\cite{dresselhaus1998physical} The PtN$_2$ nanotubes with odd $N$ values are semiconducting with indirect PBE bandgaps of are 1.24, 0.87, 0.54, and 0.40 eV,  for $N$ = 3, 5, 7, and 9, respectively. By contrast, the PtN$_2$ nanotubes with even $N$ (4, 6, 8, and 10) are quasi-metallic with nearly the same tiny PBE direct band gaps of 0.07, 0.06, 0.08, and 0.08 eV, respectively. These different electronic structures of the PtN$_2$ nanotubes with $N$ being odd and even may be because of their different symmetries. Due to the intense computational cost, we are able to calculate the HSE06 electronic structures for only two PtN$_2$ nanotubes ($N$ = 3 and 4). We find that the HSE06 bandgaps of the PtN$_2$ with $N$ = 3 and 4 are 1.96 and 0.77 eV, respectively. Note that the PBE functional once again is inaccurate to describe the bandgaps of PtN$_2$ nanotubes. This deficiency is worse for the nanotubes with even $N$ values. Assuming the trend of the PBE band gaps of PtN$_2$ nanotubes holds for the HSE06 bandgaps, namely, the bandgaps of PtN$_2$ nanotubes will decrease with increasing (odd) $N$ values and the range of tunable bandgaps is between 1.11 eV for single-layer PtN$_2$ and 1.96 eV for the ($N$ = 3) PtN$_2$ nanotube. In contrast to narrow-gap CNTs and large-gap BNNTs,\cite{blase1994stability,matsuda2010definitive} the wide range of tunable bandgaps are within the visible light spectrum, making PtN$_2$ nanotubes promising 1D materials for optoelectronics applications. 
%----------------------------------------------------------------------
\section{Conclusions}
In summary, by increasing the dimension of single-layer PtN$_2$, we have predicted a more stable structure of bulk PtN$_2$ with tetragonal AA-stacked layered structure using DFT calculations. This structure is energetically more favorable than the pyrite structure or single-layer PtN$_2$, therefore resulting in a physically positive formation energy of the single-layer PtN$_2$, which is otherwise negative if using the energy of the pyrite structure as the reference. Owing to the layered structure, our predicted bulk structure provides a promising source for mechanically exfoliated single-layer semiconducting PtN$_2$, consisting of a pattern of type 2 pentagons. We also find that applying external pressure can lead to the phase transition between the pyrite and tetragonal layered structures of PtN$_2$ and the transition pressures are about 20 and 15 GPa determined by the PBE and DFT-D3 methods, respectively. On the other hand, by reducing the dimension, we have predicted PtN$_2$ nanotubes with tunable band gaps (by varying their sizes) within the visible light spectrum. Furthermore, wrapping single-layer PtN$_2$ into nanotubes costs a comparable or smaller amount of energy in comparison to wrapping single-layer graphene and boron nitride into CNTs and BNNTs, respectively. The predicted PtN$_2$ nanotubes may find applications in optoelectronics devices. 
%----------------------------------------------------------------------
\begin{acknowledgments}
We thank the start-up funds from Arizona State University. S. Lakamsani thanks the SCience and ENgineering Experience (SCENE) program and C. Price thanks the Fulton Undergraduate Research Initiative (FURI). This research used the computational resources of the Agave computer cluster at Arizona State University. 
\end{acknowledgments}
\bibliography{references}
\end{document}